# Analyse des rôles dans les communautés virtuelles : définitions et premières expérimentations sur IMDb

**Alberto Lumbreras[1], James Lanagan[1], Julien Velcin[2], Bertrand Jouve[2]**

*1. Technicolor*
*975, avenue des Champs Blancs, 17616 Cesson Sévigné, France*
*alberto.lumbreras@technicolor.com*
*james.lanagan@technicolor.com*

*2. Laboratoire ERIC, Université Lyon 2*
*5, avenue Pierre Mendès France, 69676 Bron, France*
*julien.velcin@univ-lyon2.fr*
*bertrand.jouve@univ-lyon2.fr*

*RESUME. Analyser les rôles dans les communautés virtuelles nous permet de mieux comprendre, voire de prédire, le comportement individuel des internautes. Bien que de nombreuses approches aient été proposées, on constate un manque de généralisation des méthodes existantes et des résultats obtenus. Dans ce papier, nous passons en revue quelques théories développées à propos des rôles sociaux et nous cherchons une définition compatible à une automatisation par les machines de la détection des rôles joués par les individus dans des fils de discussions sur internet. Nous analysons ensuite le site Web IMDb afin d'illustrer notre discours.*

*ABSTRACT. Role analysis in online communities allows us to understand and predict users behavior. Though several approaches have been followed, there is still lack of generalization of their methods and their results. In this paper, we discuss about the ground theory of roles and search for a consistent and computable definition that allows the automatic detection of roles played by users in forum threads on the internet. We analyze the web site IMDb to illustrate the discussion.*

## 1. Introduction

L'analyse de rôles des individus dans les communautés sociales pose des questions sur lesquelles s'interrogent les sociologues depuis le début du XX$^{ème}$ siècle. Depuis quelques années, l'émergence de communautés virtuelles en ligne ouvre la voie à de nouvelles dynamiques et structures qui nécessitent de revisiter ces questions en sollicitant, de manière complémentaires, les connaissances et compétences issues de la sociologie et de l'informatique.

Dans ce papier, nous souhaitons répondre aux questions suivantes. Au sein de communautés virtuelles, et en particulier de fils de discussions sur internet, peut-on spécifier les différents rôles joués par les internautes, et caractériser d'éventuelles trajectoires d'individus au travers de ces rôles (Cardon *et al.*, 2011) ? Peut-on automatiser ces processus ? Afin de répondre à ces questions, nous avons besoin d'une définition de la notion de rôle qui soit utilisable par une machine et adaptée à l'analyse de ces communautés virtuelles : arrivée et départ dynamique des internautes, conversation structurée en forum, prise en compte du contenu textuel, *etc*. Une fois cette définition établie, notre démarche consiste à chercher des méthodes et des stratégies afin de réussir à capturer et analyser les rôles et leur dynamique dans une communauté virtuelle. L'une des originalités de notre approche consiste à étudier les rôles en ligne en conservant un lien théorique avec la sociologie. Nous pensons que c'est un élément important pour la clarification de ce concept protéiforme de rôle dans le contexte de l'analyse des réseaux et médias sociaux, mais cela permet surtout d'avoir une compréhension plus profonde de la dynamique des utilisateurs et des communautés. Garder en vue cette base théorique nous garantit une certaine cohérence avec l'ensemble des études existant dans la littérature, à la fois en sociologie et en informatique.

Notre objectif dans cet article est double. Il s'agit, d'une part, d'enrichir la discussion sur la définition des rôles dans les communautés en ligne. D'autre part, nous souhaitons réaliser une première analyse descriptive de l'une de ces communautés (IMDb[1]) afin de mettre en lumière sa dynamique et, surtout, souligner les défis que nous devrons affronter dans notre analyse des rôles sur ce type de forum en ligne.

La suite de cet article est organisée de la façon suivante. La section 2 ouvre la discussion sur la manière dont le rôle est défini, en particulier dans la littérature en sociologie, et ce dans l'optique d'automatiser les traitements. La section 3 donne un aperçu des différentes approches qui peuvent être employées pour détecter les rôles dans les communautés virtuelles. Avant de conclure, la section 4 propose d'étudier la communauté attachée au site IMDb afin d'illustrer la discussion.

## 2. Vers une définition des rôles pour les communautés en ligne.

Le concept de rôle a été largement discuté par les sociologues. Bien qu'il n'existe pas de consensus sur sa définition (cf. Biddle, 1986), il est clair que le rôle est associé au comportement social d'une personne, ou à son activité sociale. Du point de vue du comportement social, savoir qu'un individu est dans un rôle nous aide à comprendre la capacité de celui-ci à réaliser une action (*performance*) en fonction d'une situation donnée (Goffman, 1958 ; Parsons, 1951) ; il s'agit par exemple de l'interaction entre un docteur et son patient. Du point de vue de l'activité sociale, les rôles peuvent être utilisés afin de comprendre les jeux joués par les individus lorsqu'on analyse, par exemple, la division du travail. En théorie sociale du rôle, existent deux modèles importants: l'un macroscopique, celui des structuralistes, qui considèrent qu'un rôle est porté par la position d'un individu dans la structure sociale, l'autre microscopique, celui des interactionnistes, pour lesquels ce sont les agents et leurs interactions qui génèrent les rôles ainsi que les structures sociales. Tandis que les structuralistes mettent l'accent sur la structure des relations qui unissent les individus, les interactionnistes basent leurs analyses sur les significations que les individus attribuent à leurs actions.

[1] http://www.imdb.com/

Commençons par adopter une définition issue du mouvement structuraliste. Le rôle est alors un ensemble de comportements associés à une position dans une structure sociale (White *et al.*, 1976). Ce lien entre le rôle et la position est créé par les normes et par les attentes attachées à chaque position dans la société qui conditionnent ou rendent possibles ces comportements. De plus, les individus expriment aussi des désirs, et ils agissent en prenant en compte l'opinion que les autres individus auront d'eux en fonction de leurs actes. Une contribution importante des structuralistes a été le développement d'outils mathématiques permettant de formaliser les positions au sein de la structure sociale. Le concept d'équivalence structurelle (Lorrain and White, 1971) a été le point de départ de recherches fructueuses. Il est devenu assez vite évident que la notion d'équivalence structurelle était trop restrictive et successivement deux autres types d'équivalence vont être introduits : l'équivalence régulière et l'équivalence stochastique[2]. Les modèles par bloc (*blockmodels*) sont ensuite devenus une technique mathématique standard pour trouver ce type d'équivalences (pour quelques travaux remarqués, voir Wasserman and Anderson, 1987; Faust and Wasserman, 1992; Nowicki and Snijders, 2001).

Dans beaucoup de cas, les communautés en ligne ne nous permettent pas d'avoir accès à des informations privées, pourtant très utiles dans l'analyse des rôles, comme la liste de nos relations amicales. Cependant, il nous est possible d'observer les interactions entre les internautes. Prenons le cas d'un forum de discussion, par exemple. Lorsque des utilisateurs conversent, ils échangent des messages (*posts*) caractérisés par une étiquette temporelle et un contenu textuel. On peut alors s'attendre à trouver des motifs, des éléments de la conversation qui révèlent des

---

[2] Deux acteurs sont structurellement équivalents s'ils sont liés de la même manière à un même autre ensemble d'acteurs (autrement dit, ils partagent le même voisinage). Les équivalences régulières et stochastiques étendent cette définition en considérant les relations avec le même *type* d'acteurs et en considérant des distributions de probabilité pour caractériser ces relations.

informations sur le rôle dans lequel est chaque utilisateur (voir par exemple Welser *et al.,* 2007; McCallum *et al.,* 2007). Dans le cadre des forums de discussion en ligne, nous pouvons alors enrichir la notion d'équivalence stochastique de ces éléments interactionnistes :

*Définition* : Deux individus sont en équivalence stochastique par rapport à leurs conversations s'ils tendent à avoir le même type de conversation avec le même type d'individus. Nous dirons alors que deux individus sont dans un même rôle s'ils sont en équivalence stochastique.

La différence principale avec les définitions précédentes est que nous remplaçons le concept de relation (ex. : amitié) par celui d'interaction (ex. : conversation), objet caractérisé par un contenu (textuel) et une étendue temporelle. Nous proposons, dans ce qui suit, d'avancer la réflexion en améliorant notre compréhension des données. Avant cela, nous présentons quelques différentes utilisables pour la détection de rôles.

## 3. Eléments de stratégies pour la détection des rôles

On peut suivre différentes stratégies afin de découvrir les rôles associés aux différents individus qui constituent une communauté.

Une première approche, qui mérite à être soulignée, est dite « ethnographique ». Le sociologue étudie la communauté de l'intérieur en devenant un de ses acteurs. Dans les communautés virtuelles, l'étude ethnographique de Golder et Donath (2004) sur les forums de Usenet est probablement l'une des plus remarquables.

Dans l'analyse des réseaux et médias sociaux mettant en œuvre des méthodes informatiques automatiques, on adopte deux types de stratégies (voir par exemple, Forestier *et al.*, 2012). Les méthodes *top-down* cherchent au sein de la communauté des individus qui jouent des rôles déjà connus, prédéterminés. C'est une tâche qui peut s'avérer difficile, que de trouver des personnes qui exercent un rôle donné ou d'approfondir la compréhension que l'on a du rôle. Ces techniques ont l'inconvénient d'empêcher la découverte de rôles inattendus, émergents, que l'on pourrait observer dans la communauté. Les méthodes *bottom-up* cherchent au contraire à faire émerger des rôles sans définition *a priori*. Pour cela, il s'agit de lister un ensemble d'attributs remarquables puis d'appliquer des méthodes de classification non supervisée (par exemple de clustering à base de centroïdes, voir Anokhin *et al.*, 2012) de modèles par blocs (White *et al.*, 1976), *etc*. Les différentes catégories d'utilisateurs découvertes par ces algorithmes sont alors des candidats pour représenter des rôles différents. En particulier, les modèles par blocs ont donné de très bons résultats et sont devenus une méthode privilégiée par les sociologues.

Du fait de la grande complexité et du bruit des données issues des communautés en ligne, il nous semble cependant raisonnable d'adopter une approche de modélisation stochastique plutôt que déterministe (pour une justification détaillée, voir Robins *et al.,* 2007). Dans le contexte de l'analyse des rôles, la première tentative d'utilisation de modèles statistiques est celle de Holland *et al.* (1983).

Holland a proposé d'utiliser le modèle *p1* pour réaliser une modélisation par blocs. Cependant, réaliser une bonne modélisation n'est pas triviale et un modèle construit sur des hypothèses erronées peut amener à tirer de mauvaises conclusions. Pour ces raisons, Nowicki et Snijders (2001) ont introduit des méthodes d'estimation non paramétriques des modèles par blocs. Bien que des améliorations et des alternatives aient été proposées dans la littérature (Kemp *et al.*, 2004; Xu *et al.*, 2006; Fu *et al.*, 2009; Rodriguez, 2012, Aicher *et al.*, 2013, DuBois *et al.,* 2013), aucune ne semble avoir amélioré le modèle proposé par Snijders en sociologie. De surcroît, nous avons besoin de modéliser non seulement les interactions mais également leur contenu (voir par exemple McCallum *et al.,* 2007).

L'objectif de la prochaine section consiste à obtenir une meilleure compréhension des motifs et de la dynamique des interactions sur le site d'IMDb. Mettre en exergue les caractéristiques de ce réseau d'interaction nous aidera à mettre en place, par la suite, un mécanisme de détection automatique des rôles.

## 4. Les forums d'IMDb : un exemple de communauté en ligne

Nous définissons une communauté comme un ensemble d'individus qui échangent dans un certain contexte (ex. : un fil de discussion), partagent une culture commune (ex.: ne pas écrire en lettres capitales, ou être accueillant envers les nouveaux membres) ainsi qu'une connaissance commune, parfois partagée, du reste de la communauté (ex : qui est l'administrateur ou quels sont parmi mes amis ceux qui ont la meilleure réputation). L'existence d'une communauté requiert une masse critique d'individus en interaction avec une certaine stabilité quant à sa composition (ex.: si tous les internautes envoient un unique premier message avant de quitter définitivement la communauté sans laisser d'adresse, il semble difficile à une culture ou une connaissance commune d'émerger). Toute communauté a une structure sociale[3] qui est le niveau où les rôles entrent en action. En accord avec le caractère stochastique de notre définition des rôles, les communautés auxquelles appartient un individu sont des éléments essentiels à la spécification des rôles puisqu'elles rassemblent une partie importante des interactions de l'individu. Ainsi, avant de chercher à trouver les rôles joués par les individus, il convient d'analyser la structure des communautés sous-jacentes.

Nous analysons ici les communautés d'IMDb. L'objectif de cette section consiste à mener une analyse exploratoire afin de mieux connaître les communautés (ou la communauté, nous ne le savons pas) d'IMDb, et surtout utiliser les observations réalisées afin de nous aider à résoudre les questions mentionnées au début de cet article. IMDb est un site Web qui permet aux internautes de poster des commentaires sur des films, des séries télévisées et des documentaires. Chaque œuvre donne lieu à des fils de discussion composés de messages écrits par les internautes, comme dans le cas des forums. Notre jeu de données est composé des 250 films les mieux notés par les internautes sur le site, par les 100 films les moins bien notés, et enfin par les films faisant partie des 1000 mieux notés et qui comptabilisent également plus de 75 000 votes. Le jeu de données comprend les fils de discussion de chaque film, avec l'indication de l'auteur de chaque message. Cela représente un total de 706 films, 102 000 fils de discussion, 1 115 616 messages et 117 000 internautes. Ces données ont été téléchargées en septembre 2012.

[3] Selon Nadel (1957), la structure sociale est « le réseau des relations obtenus entre des acteurs à partir de leur capacité à jouer un rôle les uns vis-à-vis des autres ».

### *1.1. Analyse descriptive des catégories données a priori*

Après avoir quelque peu « joué » avec les données, nous avons choisi de diviser les utilisateurs en cinq catégories assez simples, dans l'objectif de mieux comprendre le comportement des internautes sur IMDb. Remarquons qu'il ne s'agit pas là de rôles, mais d'une classification simple des individus. Avec cette analyse segmentée, nous essayons d'éviter une sorte de « cécité statistique » qui serait causée par l'agrégation erronées d'individus aux comportements très différents : *1-posters* (47.5%): les internautes qui ont écrit un seul message (*post*) ; *1-threaders* (5.3%): les internautes qui ont participé à un seul fil de discussion (*thread*) ; *réguliers* (37.5%): les internautes qui ont contribué à entre 2 et 10 fils ; *pros* (9%): les internautes qui ont contribué à entre 10 et 100 fils ; *super actifs* (0.7%): ceux qui ont contribué à plus de 100 fils.

*Tableau 1. Attributs d'activité pour chaque catégorie (médianes)*
*(Vie : nombre d'heures entre le premier et le dernier message de l'internaute dans le fil. Naissance : nombre d'heures entre la création du fil et le premier message posté par l'internaute).*

| | Films | Fils | Posts | Taille du fil | Posts/ fil | Mots/ post | Vie/ fil | Naissance/ fil |
|---|---|---|---|---|---|---|---|---|
| 1-posters | 1 | 1 | 1 | 31 | 1 | 36 | 0 | 55 |
| 1-threaders | 1 | 1 | 2 | 31 | 2 | 50 | 2.8 | 1.7 |
| Regulars | 2 | 3 | 3 | 31 | 1 | 46 | 0 | 106 |
| Pros | 9 | 18 | 24 | 29 | 1.4 | 52 | 0.84 | 107 |
| Super actifs | 15 | 160 | 274 | 25 | 1.53 | 50 | 0.6 | 24 |

*Tableau 2. Attributs sociaux dans les fils de discussion pour chaque catégorie (médianes)*
*('in' et 'out' indiquent le degré entrant (réponses reçues) et le degré sortant (réponses données)).*

| | In/ fil | Out/ fil | (In/Out)/ fil | % Fils initiés | % Posts ignorés (initient fil) | % Posts ignorés (n'initient pas fil) |
|---|---|---|---|---|---|---|
| 1-posters | 1 | 1 | 0 | 16% | 78% | 48% |
| 1-threaders | 2 | 2 | 1 | 41% | 35% | 37% |
| Regulars | 0.5 | 1 | 1.2 | 18% | 61% | 41% |
| Pros | 1.23 | 1.27 | 150k | 16% | 57% | 37% |
| Super actifs | 0.78 | 1.34 | 76k | 11% | 50% | 36% |

Nous résumons ci-dessous les principales observations réalisées à la lumière de cette première catégorisation.

*1-posters*

Les 1-posters sont des internautes qui ont écrit un seul message dans IMDb. Le tableau 2 montre que, à l'intérieur des fils de discussion dans lesquels un 1-poster participe, 16% de ces fils ont été commencés par lui. Ce pourcentage est similaire pour la plupart des catégories d'internautes. Cependant, le taux d'ignorance, c'est-à-dire la probabilité qu'aucun internaute ne réponde à ces messages, est considérablement plus haut : personne ne répond à 78% des fils débutés par un 1-poster. Il semble qu'ils ont du mal à engager d'autres internautes dans leurs conversations. Si, par contre, ils participent à un fil qui a déjà été commencé, leur taux d'ignorance baisse à 48%.

*1-threaders*

Les 1-threaders ont écrit plusieurs messages, mais toujours dans un seul fil de discussion. Leur activité est assez étonnante. Si nous définissons la vie d'un utilisateur dans un fil de discussion comme le nombre d'heures entre son premier et son dernier message, la vie des 1-threaders est la plus longue de toutes (2,8h). Leur date de naissance, définie comme l'heure calculée entre la création du fil de discussion et le premier message posté, est particulièrement faible (1,7h). Cela s'explique probablement par le fait qu'ils ont tendance à commencer de nouveaux fils de discussion. Et lorsque cela arrive, ils obtiennent la plupart du temps une réponse avec un taux d'ignorance inferieur (35%) à celui des autres groupes.

*Réguliers*

La particularité la plus remarquable des réguliers est que ces internautes entrent en général tardivement dans la conversation (106 heures après le début du fil de discussion). D'ailleurs, ils écrivent typiquement un seul message dans les fils de discussion.

*Pros*

La caractéristique des pros est leur ratio élevé de *in-degree/outdegree*[4] : ils reçoivent beaucoup plus de réponses qu'ils ne postent eux-mêmes de messages. Si l'on considère que les réguliers ont un ratio de 1.2, et les pros un ratio de 150 000, il semble qu'il peut exister un point d'inflexion où ce ratio augmente rapidement**.** Il s'agit probablement d'internautes cinéphiles qui connaissent très bien la culture de la communauté.

[4] Le cas où l'utilisateur ne répond à aucun post dans le fil (Out nul) provoque une division par zéro. Afin de prendre en compte ces cas dans le calcul, nous définissons par convention In/Out à $10^6$.

*Super actifs*

Bien que les super actifs ne commencent pas autant de fils de discussion que les réguliers ou que les pros, ils entrent en moyenne plus tôt dans la conversation. Leur taux d'ignorance est inférieur à celui des autres groupes. Par contre, ils ont un *in-degree*/*out-degree* encore supérieur à celui des pros.

Cette première analyse confirme que, comme dans d'autres forums, il existe bien différents types d'internautes sur IMDb. La dynamique et la structure des conversations dans les fils dépendent bien sûr de la nature de la conversation en elle-même, mais aussi du type d'individus qui y participent. Une analyse en termes de rôles permettra de mieux comprendre comment et pourquoi les conversations se développent d'une manière ou d'une autre. De plus, les forums eux-mêmes imposent certaines contraintes sur le type de structure qui peut émerger. Contrairement à d'autres structures sociales, les forums gravitent le plus souvent autour d'une question initiale posée par un internaute. On observe que la catégorie, et probablement le rôle, de la personne qui pose la question va déterminer en grande partie le succès rencontré par le fil de discussion. Le rôle a une incidence sur la structure, et nous devons prendre cette observation en compte dans notre future modélisation.

### *4.2. Etude des motifs dans les fils de discussion*

Les motifs sont des interconnexions dans les réseaux complexes dont la fréquence d'apparition est significativement plus importante que celles qui apparaîtraient dans un réseau similaire mais aléatoire (Milo, 2002). Dans l'analyse des réseaux sociaux, l'analyse de motifs nous permet de découvrir des schémas récurrents dans les conversations (Adamic, 2008 ; Qazvinian, 2011). La figure 1 montre tous les motifs possibles avec 3 sommets. Nous avons analysé ce type de motifs dans les fils de discussion car il s'agit d'une manière commode pour commencer à comprendre la structure des conversations. En utilisant FANMOD (Wernicke *et al*, 2006), nous calculons automatiquement la distribution des 3-motifs dans chaque fil de notre collection.

*Figure 1. Motifs composés de 3 sommets*

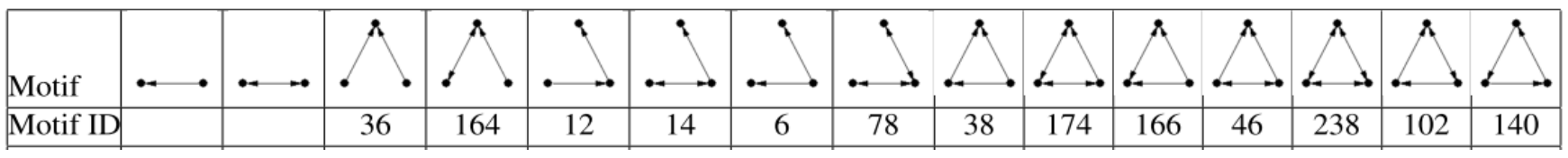

| Motif | | | | | | | | | | | | | | |
|---|---|---|---|---|---|---|---|---|---|---|---|---|---|---|
| Motif ID | | | 36 | 164 | 12 | 14 | 6 | 78 | 38 | 174 | 166 | 46 | 238 | 102 | 140 |

La figure 2 permet de visualiser l'occurrence de ces motifs pour différentes longueurs des fils de discussion, ainsi que leur z-score. Le z-score est défini comme la différence entre la fréquence d'un motif dans le réseau original et sa fréquence moyenne dans un ensemble de réseaux générés aléatoirement (« randomisés ») – à partir d'une distribution des liens du réseau original.

On peut observer une phase de transition entre fils de discussion courts et longs située autour de 100 messages. Passée cette longueur, la population des motifs atteint une phase de stabilité. Nous détaillons ci-dessous les principaux motifs observés.

*Figure 2. Relation des motifs avec la longueur des fils de discussion.*

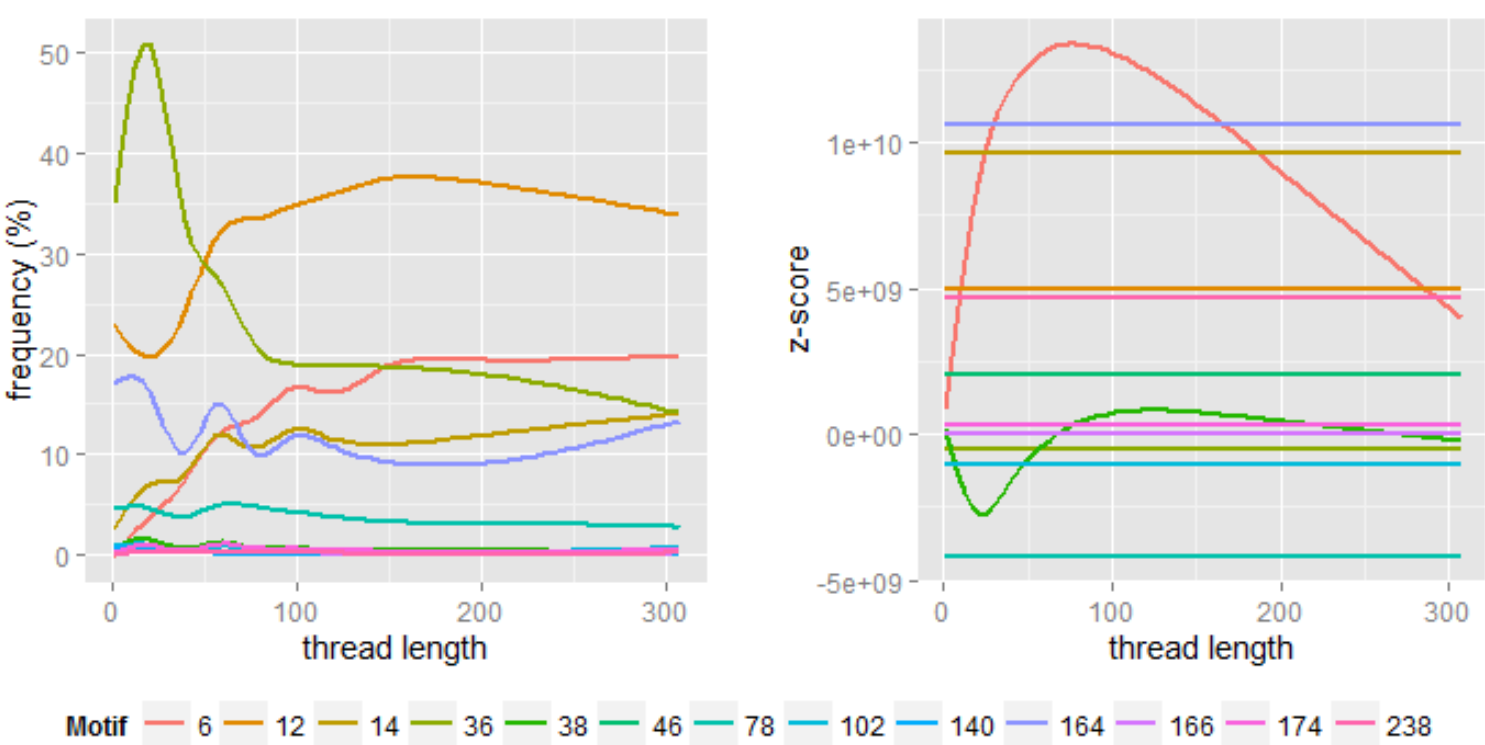

Le motif 36 est le motif qui domine dans les fils de discussion courts. Ce motif est quelquefois associé à un schéma de type Question-Réponse ; il s'agit par exemple des forums où un internaute pose une question et on lui répond sans qu'il y ait d'interaction. Tandis que ce motif est prédominant dans tous les fils de discussion de faible étendue, son importance relative diminue dans les fils de discussion plus longs. Le motif 12 prend la place du motif 36 dans les fils longs. Ce motif indique une conversation en forme de cascade : un internaute répond à un message, et on lui répond à son tour, etc. Le motif 164 correspond également à un schéma de type Question-Réponse, mais où le questionneur répond à son tour à l'une des réponses. On peut voir ce schéma comme le niveau minimum pour assister à un embryon de conversation. Il est intéressant d'observer que ces motifs sont également présents dans les conversations courtes. Les motifs 14 et 6 n'existent pas dans les fils de discussion courts, mais ils ont une présence significative dans les fils de discussion plus larges. Le motif 6 correspond à un utilisateur qui répond à la question originale, mais aussi à un autre internaute dans le forum. Le motif 14 est similaire, mais avec une interaction bidirectionnelle, ce qui signifie que ces deux internautes ont bien eu une petite conversation. Le motif 78 est le dernier des motifs avec une fréquence importante. Il correspond à des internautes qui posent une question et qui répondent ensuite à plusieurs internautes qui lui ont répondu.

Même s'il est important de connaître la proportion de ces motifs dans les conversations, il est possible que leur présence soit provoquée par la topologie des fils de discussion et non par la nature de la conversation.-La figure 2 montre que le z-score pour la plupart de ces motifs principaux est assez élevé. Cependant, la présence du motif 36 (type Question-Réponse) est associée à un z-score proche de zéro. Cela implique que ce motif aurait une présence similaire à celui d'un réseau aléatoire. D'un autre côté, le motif 78 est beaucoup plus fréquent que ce qu'on pouvait attendre.

Bien que les réseaux aléatoires aient été générés à partir des fils de discussion originaux, nous devons prendre des précautions dans l'interprétation des résultats car la topologie d'un fil de discussion est très probablement générée une fois que l'objectif du forum est bien établi. Par exemple, lorsqu'un internaute commence un nouveau fil de discussion avec une question à propos d'un film, cela conditionne nécessairement la forme que va prendre la discussion (qui va certainement commencer dans un style de Question-Réponse). On peut donc s'attendre à ce que la randomisation de ce fil de discussion implique un biais vers le motif 36.

L'analyse que nous venons de faire apporte un premier éclairage sur le type de motifs sur lesquels reposent ces forums. Une manière classique de modéliser des réseaux sociaux est basée sur des modèles qui génèrent des réseaux en supposant des tendances comme la réciprocité ou la transitivité (voir Robins *et al.,* 2007). A la lumière de notre analyse, il semble que la réciprocité soit une caractéristique importante des forums. Néanmoins, dans les fils les plus courts, la transitivité (motif 38) est moins fréquente que ce qui est attendu. Mise à part la réciprocité, les motifs 6 and 12 (cascade) semblent être de bons candidats pour intégrer un modèle. On remarque, aussi, que la fréquence de certains motifs change entre les fils courts et les fils plus longs. Ainsi, les modèles devraient prendre en compte ces phénomènes si, comme nous le pensons, ce comportement n'est pas expliqué de manière intrinsèque par l'évolution naturelle d'une population d'individus plus grande. Enfin, parcourant ces forums, nous avons souvent observé que l'évolution des discussions dépend beaucoup du contenu de la contribution, et même de la question initiale. Il sera important d'étudier la part que prend le contenu dans l'évolution d'une discussion par rapport à l'effet que peut avoir la structure générée dans cette même évolution.

## 5. Conclusion

Dans cet article, nous avons initié une discussion sur le concept de rôle, à la fois sur la base des définitions héritées de la sociologie mais dans l'optique de mieux analyser les rôles joués par les internautes dans les communautés virtuelles. Ainsi, nous proposons de combiner les points de vue structuraliste et interactionniste afin de préciser et formaliser les rôles au sein de ces communautés. Nous proposons une définition stochastique de l'équivalence de rôle basée à la fois sur la morphologie du réseau d'interactions et sur la nature (contenu) des conversations entre individus. Ensuite, nous avons réalisé une première analyse exploratoire de la communauté associée au site IMDb, ce qui nous a permis de trouver certains motifs de comportement dans la discussion. L'observation d'un groupe si important d'individus ayant écrit un unique message pose la question de la trajectoire de vie des internautes. Quelles sont les circonstances qui font qu'un nouvel internaute abandonne la communauté après avoir écrit un message ou, au contraire, devienne un utilisateur engagé ? Existe-t-il une propension au succès ou à l'abandon ? Cela pourrait dépendre d'attributs non observés (ex. : timidité, cinéphilie, etc.), mais aussi de son expérience avec les autres individus (ex. : message systématiquement ignoré). De plus, nous avons analysé les motifs conversationnels et, contrairement à d'autres types de réseau, nous n'y avons pas trouvé de transitivité dans les fils courts (moins de 50 messages), tandis que d'autres motifs dominent. Cela nous incite à éviter certains modèles existants dont les hypothèses ne seraient pas compatibles avec ces forums. Une étude plus détaillée nous permettrait de discerner les éléments fondamentaux du forum qui rendent possible l'émergence de la discussion.

Concernant la dynamique des discussions, l'analyse des différents motifs observés permet-elle de renseigner sur l'influence des rôles dans cette dynamique ? On peut penser que la composition initiale des motifs dans un fil de discussion est déterminante, donc prédictive, dans l'évolution future de ce fil.

La sociologie a souvent analysé les communautés en termes de centre-périphérie. Peut-être existe-t-il des internautes constants qui forment un noyau dans la structure sociale ? Ce noyau permettrait à la structure de se maintenir tandis qu'une périphérie d'internautes aux rôles plus éphémères se renouvelle constamment.

De manière générale, il ne s'agira pas uniquement de trouver des rôles, mais bien de comprendre ces rôles, leur genèse, leur dynamique, leur relation avec les communautés dont ils sont issus.

## Références